# FUSING TEXT AND IMAGE FOR EVENT DETECTION IN TWITTER

Samar M. Alqhtani, Suhuai Luo and Brian Regan

School of Design, Communication and IT, The University of Newcastle, Callaghan NSW 2308, Australia

## ABSTRACT

*In this contribution, we develop an accurate and effective event detection method to detect events from a Twitter stream, which uses visual and textual information to improve the performance of the mining process. The method monitors a Twitter stream to pick up tweets having texts and images and stores them into a database. This is followed by applying a mining algorithm to detect an event. The procedure starts with detecting events based on text only by using the feature of the bag-of-words which is calculated using the term frequency-inverse document frequency (TF-IDF) method. Then it detects the event based on image only by using visual features including histogram of oriented gradients (HOG) descriptors, grey-level co-occurrence matrix (GLCM), and color histogram. K nearest neighbours (Knn) classification is used in the detection. The final decision of the event detection is made based on the reliabilities of text only detection and image only detection. The experiment result showed that the proposed method achieved high accuracy of 0.94, comparing with 0.89 with texts only, and 0.86 with images only*.

## KEYWORDS

## 1. INTRODUCTION

The world has greatly changed, including the way people communicate. One of the most recent phenomenon have been the social media collections that are available over the Internet. Social media is simply defined as applications and websites that allow the users to create and share information or content, as well as to participate in other activities such as social networking[1]. They allow people to interact, exchange information concerning their lives such as uploading photos of events and current issues going on in their lives. Today, it is not only used by people for personal purposes, but also by organizations for corporate issues.

The growth of social media over the last one decade has been tremendous with numbers of people joining doubling almost on a daily basis. This growth has brought about the need for scalable, robust and effective techniques of managing, as well as indexing the content they produce. Anything going on in the world is shared and communicated through the Internet, especially in social media. Social media offers people the chance to interact, comment on events, and send instant messages all over the globe without geographical barriers. They include Facebook, YouTube and Twitter amongst others. The social media platforms have opened up many research opportunities because of the amount of information they possess. This information can be used for many purposes including things such as prediction and detection of events and even as warning systems. Events are one of the most important indications of people's memories. They are a natural way through which people refer to any observable occurrences that bring people together in the same places and time to undertake similar activities [2]. They are quite useful in making sense of the world around us, as well as in helping people recall the experiences they have gone through. Social events on the other hand refer to those events that are attended by people and presented in the multimedia content that is shared through online websites. Such events can include disasters, concerts, sporting events, public celebrations and protests amongst others.

This paper is about detecting events based on information collected from social media. Specifically, weuse Twitter, which is one of the social media applications that has rapidly emerged over the last few years. Many people use Twitter for reporting events as they happen in the real world [3]. This social media currently has over 500 million registered accounts all over the world that generate about 340 million messages daily. Most of these messages contain personal updates, opinion concerning current issues, moods, genera life observations and events amongst others. The readily available wide range of data from Twitter offers an ideal source for mining information for this research. There have been proposals concerning event mining that make use of the Tweet texts. However, none of them has proposed the use of images in the data mining, which makes the analysis solely textual. In this paper, we seek to use both textual and visual mining to detect events in order to improve the performance of the event detection considering the retrieved results will depend on the amount of information collected [4].

The paper aims to develop an accurate and effective detection method to detect events from the Twitter stream. We monitor the Twitter stream in order to pick up texts that have photos, which are then stored in a database. It is followed by an extraction of features in both text and photos to be applied in the mining stage. It uses "bag of words" as the features of the text which will be collected using the Term Frequency-Inverse Document Frequency method (TF-IDF) [5]. For visual features, it uses Histogram of Oriented Gradients (HOG) descriptors for object detection, Grey-Level Co-occurrence Matrix (GLCM) for texture description, and color histogram[6-8]. Support Vector Machine (SVM) classification algorithm will be applied to mine the data and get the accuracy measure[9].

The success of the proposal will result in breakthrough in event detection based in social multimedia data, and make the mining result more effective and accurate. This paper is organized as follow: in the next section some researches and progress in the area of event detection based in social multimedia data are presented, followed by the proposed method in the third section. In the last section the experiments and evaluation of event detection are described.

## 2. EVENT DETECTION BASED ON SOCIAL MULTIMEDIA DATA

The multimedia facet of social networking sites significantly increases engagement of the users. Multimedia is content and media that employs content forms and includes a combination of images, texts, audio, video, and interactivity content forms [10]. Generating high quality video and photo content for social media has become a key interest area in social media. Social media sites include Facebook, Google Plus+, Twitter, Linked In, Tagged, Pinterest, Instagram, Tumblr, and Flickr. All social media sites allow for publishing as well as sharing of multimedia content which plays an important role in building of relationship between users and content and among users, For example, a user can share with friends and comment on a YouTube video on Facebook. The large the number of users posting multimedia content, for instance, posting pictures, videos, updates and texts, the more successful the participatory social network site.

The advance of systems, in particular the web-based social systems, has heightened in recent years in an exponential manner. Recently, academics and researchers started to examine a range of data mining techniques to assist experts enhance social media [11]. These techniques permit experts to discern novel information derived from users' application data. Lately, a range of community services as well as web-based sharing like YouTube and Flickr have made a massive and hastily mounted amount of multimedia content accessible online. Content uploaded by partakers in these vast content pools is escorted by wide-ranging forms of metadata, like descriptive textual data or social network information.

Twitter and social media trends have drastically changed in the recent past with millions of users going to the platform to chat, exchange ideas or share stories[12]. As a result, this platform has formed a rich place for news, events and information mining. However, due to the huge burst in information, data mining in Twitter is a complicated venture that requires a lot of skills and information on important ways of undertaking data mining. Twitter and social media sites have traffic overflows which are multiple and huge in terms of the frequency [13]. For instance twitter receives over 80 million tweets a day and this leads to billions of tweets per month. As a result, event prediction and detection requires the use of complex algorithms which go through the text and images in keyword matching process[14]. One of the requisite skills includes extraction features and algorithms that could be deployed in mining data such text and data. Several data mining tools and algorithms have been developed with the capabilities and purpose of analysing data and text. Researchers and other people have come up with techniques of mining data from social media with the use of different types of algorithms[15]. For instance, through the use of mining tools such as RData Mining tool, we can target some key words to mine within events and other forms of data from twitter streams. There are several techniques that could be used in the process of mining text and multimedia data in social media channels [16]. One of the important uses of data mining within social media is on event detection in twitter or social media channels. Event detection within social media through the use of different data mining techniques and algorithms is common and growing within the social media sphere [17]. Techniques such as mining events through geo-tagged events and geo-tweet photos have been utilized in regions such as Japan and Singapore to identify events such as Typhoons and floods. These techniques have been successfully in finding information on different events. These mining techniques make use of keywords to within bursts of Twitter streams for matching identities [18]. As a result, these tweets are grouped into certain databases where they are analysed. The processes of mining involved searching for keywords with emphasis on event detection with focus on words those are frequent. Then these event keywords will be unified while geo-tweet photos which correspond with the keywords will be clustered and grouped together. Each of these photos will be matched against these events and shown on the map. In the process of event detection it is imperative to look into variable factors such as distinct languages and locations. This approach will address the gap found in the process and tasks which require quick event detection in Twitter and social media circles [19].

Twitter is one the social media sites that have tremendous traffic overflows which are multiple and huge in terms of the frequency. For instance twitter receives over 80 million tweets a day and this leads to billions of tweets per month. As a result, event detection requires the use of complex algorithms which go through the text in keyword matching process [18]. For text mining in event detection by using Twitter data, there are different way to detect event like using part of speech technique, Hidden Markov Model (HMM), and Term Frequency and Inverse Document Frequency (TF_IDF).

As the saying goes a picture is worth a thousand words. Nowadays social media users find it much more convenient and enjoyable than ever before to express their opinions by posting pictures, attaching video clips rather than just typing a message. Mobile social network application developers also introduce features to allow users to take pictures and then upload them through a simple click. 3Compared with text information, multimedia contents are more eye-catching and entertaining. Social media sites are used in posting multimedia data such as photos, videos and other content which allow information sharing such as events. In the process of event detection we have to make use of algorithms and techniques that allow searching, extraction and storing of multimedia data from social media streams. Due to the huge volume of content such as videos, images and other content, we have to utilize techniques with emphasis on content-basis image retrieval algorithms [15].

This paper will illustrate data mining technique for the purpose of event detection through the use of two algorithms in the process of extraction. These methods will utilize different algorithms in extracting text and photo streams from twitter. The failure of having an accurate event detection method in the process of social media or twitter mining precipitates a problem that needs a solution [20]. As a result the use and combination of Term Frequency-Inverse Document Frequency method (TF-IDF) and for image content, Histogram of Oriented Gradients (HOG) descriptors for object detection, Grey-Level Co-occurrence Matrix (GLCM) for texture description, and color histogram might yield better results in the process of event detection. These algorithms are effective in the process of mining and obtaining information on different events.

## 3. THE PROPOSED METHOD

The aim of this paper is to develop advanced algorithms for event detection, especially for text and image in social media data. The algorithm will be used in advance to fuse different features that can be extracted from multimedia data in Twitter. It will achieve a breakthrough in event detection. The proposed framework is illustrated in Figure 1.

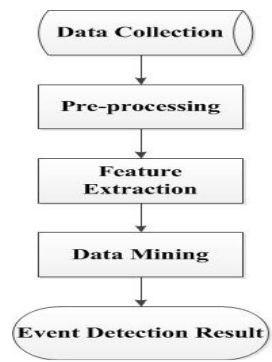

Figure 1. The Proposed Method

In this section, we explain the detail of each step of the proposed system. Before that, we monitor Twitter stream to pick up tweets having both text and photos, and store them into a database. Then, we detect the event in text data only, image data only, and fuse the image with the text in last method.

## 3.1. Text Data

Text data mining is useful for research into social media because it gives researchers the ability to automatically detect event in Twitter. We use the text data to detect event in this step and our method is depicted in figure 2.

Tweet message is written in sentence in general of which the maximum number of letters is 140. To do event detection by using text data in Twitter, we filtered out tweets that contain non –Latin characters, trying to maintain a corpus of English tweets. Although we managed to remove all East Asian tweets, our corpus still contained some non–English tweets mainly in Spanish and Dutch. Lowercase all words in the tweets. Then we follow the procedure:

a)  Tokenize

Convert the string to a list of tokens based on whitespace. This process also removes punctuation marks from the text.

b)  Stop word filtering

Eliminate the words which are common and their presence does not tell us anything about the dataset, such as: the, and, for, etc.

c)  Stemming filtering

Reduce each word to its stem, removing any prefixes or suffixes.

d)  Indexing

We index the data after filtering and stemming by using TF-IDF which is a weighting scheme that weighs features in tweets based on how often the word occurs in an individual tweet compared with how often it occurs in other tweets. Then by measuring the weight for each keyword, we can decide the related event.

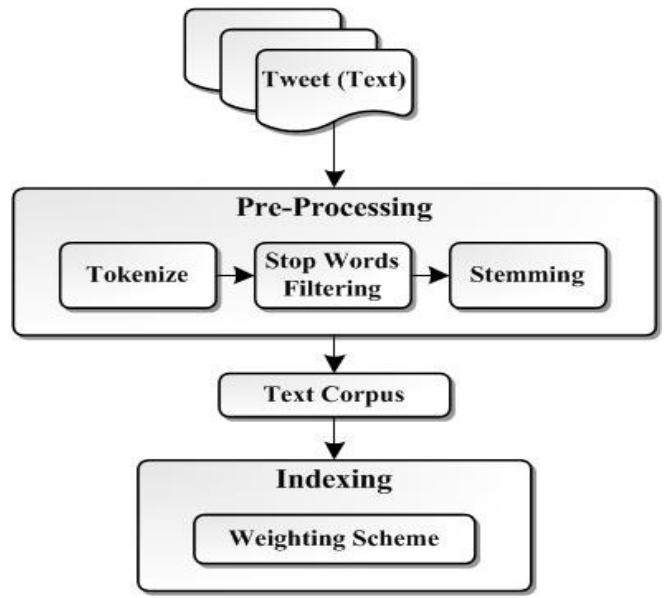

Figure 2.The block diagram of text data

## 3.2. Image Data

Image mining uses three distinguishable types of feature vectors for images description to ensure the accuracy is very high at any particular case. These feature vectors are Histogram of Oriented Gradients (HOG) descriptors for object detection, Grey-Level Co-occurrence Matrix (GLCM) for texture description, and color histogram. The method is depicted in figure 3.

HOG descriptors are used in computer vision and image processing for object detection. This technique works on the occurrences of gradient orientation in localized portions of a particular image. The object appearance and the shape within an image are described by the distribution of intensity gradients. GLCM is used for purposes of texture description such as land surface or even an extensive ocean. It is defined as the distribution that is defined over an image to be the distribution of co-occurring values at a given offset values. Its main applicability is to measure the texture of surfaces. Another aspect of feature extraction is the color histogram. This aspect is used in image processing and photography and it is defined as the representation of distribution of colours of an image.

In addition to features extraction, we build a data model by using the data mining techniques. In our work, we have used Support Vector Machine (SVM) classification which analyses input data and finally recognize patterns that are used in classification an. A Support vector machine model is the representation of the example elements as points in a defined space layout that is mapped so that the examples of the separate categories are actually divided by a distinctive gap which is wide enough.

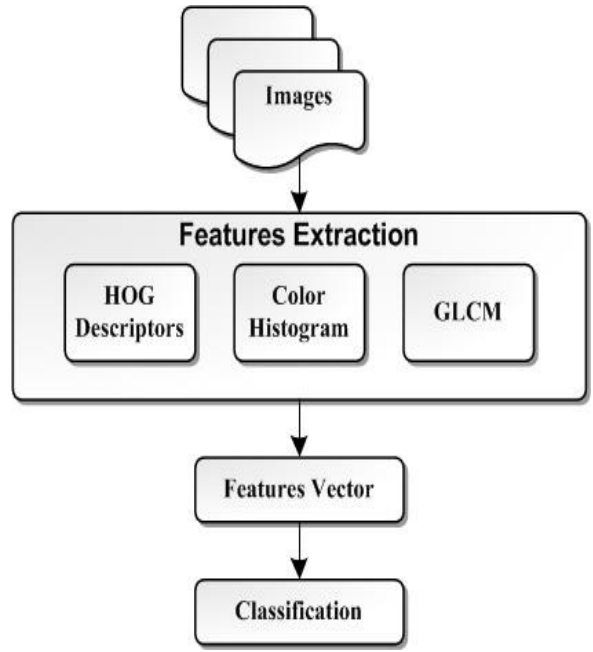

Figure 3.Theblock diagram of image data

### 3.3. Fusion Image and Text Data

For multimedia data, we apply fusion for text and image by combining text and image features. We use our database that contains both text and photos to the proposed event mining system which consists of event keyword detection, event photo classification photo selection. The features we extract in this step are the combination of features in section 3.1, and section 3.2.

In our fusion method, if the tweet text mining score is less than a threshold, this means that the text mining is not reliable so the tweet is classified using the image only; otherwise, the tweet is classified using the text.

## 4. EXPERIMENTAL RESULTS

In the experiment, we used about one million tweets which contain texts and photos posted about Indonesia Air Asia Flight 8501,which were collected from the Twitter stream in28 December 2014.We train our algorithms on our data. We divided the data into three equal parts. We use the earliest two thirds of the data as training and validation sets.

As results of event keyword extraction from our text data, we obtained 100 keywords related to found event such as: found, black box, rescue, passenger and others. Then, we produced a composite weight for each term. For image, we classify our data by using Support Vector Machine (SVM) classification into two classes depending on the event. Class 1 represents plane is found, and class 2 represents plane is not found. By preparing a training set, we can produce a model to classify tweets automatically into each class. Figure 4 represents the image's sample for the two classes.

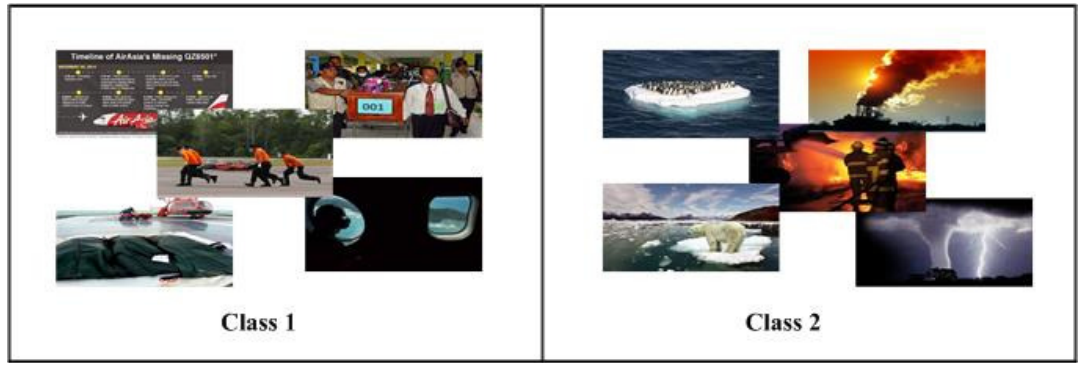

Figure 4. Some example images for both classes where class1 meansplane is found, and class 2 means plane is not found

We prepare three groups of features for each tweet to detect the event as follows:

- First group features is the text features for text mining.
- Second group features is the features for image mining.
- Third group features is the fusion for text and image features.

Finally, we measure the accuracy for each mining method by apply the following equation.

$$A = \frac{TP + TN}{TP + TN + FP + FN},$$

where A represents the accuracy for the event detection method, TP, TN, FP and FN represents true positive, true negative, false positive and false negative respectively.In our classification, airplane-crash has happened class is a true positive.

From the experiment, we find that accuracy from event detection model for the fusion of text and image gave more accurate result and made the event detection more effective. The result is shown in Table 1.

Table 1. Result for the method's accuracy

| Data Type | Text Data | Image Data | Fusion Images and Text Data |
|---|---|---|---|
| Accuracy | 0.89 | 0.86 | 0.94 |

## 5. CONCLUSION AND FUTURE WORK

In this paper, we proposed an event detection method to detect events from Twitter stream, by applying mining tool for Twitter streams that have texts and photos. It has been proved that mining both visual as well as textual information will give accurate and effective result. Our method achieves better accuracy when we fuse text and image in mining algorithm for event detection application. Future work will focus on using different method of fusion for text and image features, and adding more effective features, which makes the event detection better.

## ACKNOWLEDGEMENTS

The primary author and related research is sponsored by Najran University in Saudi Arabia.